\documentclass[10pt,conference]{IEEEtran}
\usepackage{cite}
\usepackage{amsmath,amssymb,amsfonts}
\usepackage{graphicx}
\usepackage{textcomp}
\usepackage{xcolor}
\usepackage[hyphens]{url}
\usepackage{fancyhdr}
\usepackage{hyperref}
\usepackage{algorithm}
\usepackage{algpseudocode}
\usepackage[table]{xcolor}

\newcommand{\hpcayear}{2027}

\title{Operating System Support for Accelerators as First-Class Compute Elements in Heterogeneous Systems}

\def\hpcacameraready{} 

\newcommand\hpcaauthors{Mohammad Shahidzadeh$\dagger$ and Lesley Shannon$\ddagger$}
\newcommand\hpcaaffiliation{Simon~Fraser~University, Canada$\dagger$, Simon~Fraser~University, Canada$\ddagger$}
\newcommand\hpcaemail{msa417@sfu.ca}

\newcommand*{\ToolName}{MEERKAT}

\author{
  \ifdefined\hpcacameraready
    \IEEEauthorblockN{\hpcaauthors{}}
      \IEEEauthorblockA{
        \hpcaaffiliation{} \\
        \hpcaemail{}
      }
  \else
    \IEEEauthorblockN{\normalsize{HPCA \hpcayear{} Submission
      \textbf{\#1622}} \\
      \IEEEauthorblockA{
        Confidential Draft \\
        Do NOT Distribute!!
      }
    }
  \fi 
}

\fancypagestyle{camerareadyfirstpage}{%
  \fancyhead{}
  
  \fancyhead[C]{
    \ifdefined\aeopen
    \parbox[][12mm][t]{13.5cm}{\hpcayear{} IEEE International Symposium on High-Performance Computer Architecture (HPCA)}    
    \else
      \ifdefined\aereviewed
      \parbox[][12mm][t]{13.5cm}{\hpcayear{} IEEE International Symposium on High-Performance Computer Architecture (HPCA)}
      \else
      \ifdefined\aereproduced
      \parbox[][12mm][t]{13.5cm}{\hpcayear{} IEEE International Symposium on High-Performance Computer Architecture (HPCA)}
      \else
      \parbox[][0mm][t]{13.5cm}{\hpcayear{} IEEE International Symposium on High-Performance Computer Architecture (HPCA)}
    \fi 
    \fi 
    \fi 
    \ifdefined\aeopen 
      \includegraphics[width=12mm,height=12mm]{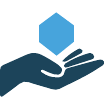}
    \fi 
    \ifdefined\aereviewed
      \includegraphics[width=12mm,height=12mm]{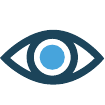}
    \fi 
    \ifdefined\aereproduced
      \includegraphics[width=12mm,height=12mm]{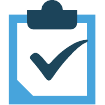}
    \fi
  }
  \fancyfoot[C]{}
}
\begin{document}
\maketitle


\newcommand{\hpcaheight}{0mm}
\ifdefined\eaopen
\renewcommand{\hpcaheight}{12mm}
\fi

\begin{abstract}
\small
Heterogeneous systems-on-chip (SoCs) increasingly rely on domain-specific hardware accelerators (HAs) to deliver performance and energy-efficiency gains beyond what general-purpose processors can provide. However, operating systems (O/S), such as Linux, assume that processors are the Compute Elements (CEs) and  HAs are peripheral devices rather than specialized CEs. As a result, HA execution incurs substantial software (SW) overheads that reduce the effective benefits of hardware (HW) acceleration.

This work demonstrates that significant inefficiencies arise from ``O/S \emph{peripheral device taxes}'' (O/S taxes), including SW stack-accelerator communication, invocation, interrupt handling, and scheduling. 
We built~\ToolName\footnote{https://github.com/mohammadshahidzade/MEERKAT}, an Field-Programmable Gate Array (FPGA)-based framework for configurable multicore systems that include HAs; it allows the user to migrate Linux O/S support from SW to HW 
and identify, study and quantify the impact.
Here we study 
four major sources of O/S tax and introduce 
techniques to mitigate them. These include a virtual-memory-aware (VMA) communication method that enables efficient HA access to application data, a lightweight HA invocation interface, Interrupt Policy Adaptation (IPA) for dynamic polling and interrupt selection, and an HA-aware scheduler that dynamically distributes work across Central Processing Units (CPUs) and multiple HA instances.

Experimental results demonstrate up to 260$\times$ faster communication and 2--6$\times$ performance improvements in wall clock time over conventional Schedulers. Furthermore, our study reveals that overall system performance depends not only on HA speed, but also on O/S overheads, resource availability, and scheduling decisions. These findings highlight the importance of treating HAs as first-class computational resources and motivate a more accelerator-centric O/S design for future heterogeneous systems.

\end{abstract}


\section{Introduction}\label{sec:intro}

With the end of Dennard scaling, heterogeneous System-on-Chips (SoCs) that integrate general-purpose processors with specialized hardware (HW) accelerators have emerged as a key approach for sustaining performance growth and improving energy efficiency \cite{esmaeilzadeh2011dark,cong2018customizable}. 
As a result, considerable research effort has been devoted to designing increasingly faster and more specialized accelerators. However, these accelerators are often developed and evaluated in isolation, with limited consideration of the applications they ultimately serve or the system-level environment in which they operate. In particular, relatively little attention has been paid to how Operating System (O/S) abstractions, resource management policies, and software (SW) interfaces influence accelerator utilization and overall application performance \cite{kurth2017hero,genc2021gemmini}.

Despite their growing importance, accelerators continue to be treated as peripheral devices within most O/Ss. Modern O/Ss were designed around a CPU-centric execution model in which Central Processing Unit (CPU) cores are the primary Compute Elements CE responsible for scheduling, resource management, and execution. Accelerators, in contrast, are typically accessed through device-driver interfaces and remain outside O/S abstractions\cite{ismail2011fuse}. As accelerators increasingly perform a substantial fraction of application computation, We argue that accelerators should increasingly be treated as first-class CE rather than passive devices\cite{suresh2024mozart}.

The lack of first-class O/S support introduces a variety of overheads that reduce the effective benefits of HW acceleration\cite{genc2021gemmini,kurth2017hero}. 
These overheads can significantly impact application performance, particularly for fine-grained accelerators and latency-sensitive workloads. In this work, we refer to such overheads as ``O/S \emph{peripheral device taxes}'' (O/S taxes). 
There are four major sources of overhead in tightly integrated heterogeneous SoCs: communication, invocation, interrupt handling, and scheduling overheads.

One major source of O/S tax arises from communication between SW and HW accelerators \cite{cota2015analysis,koenig2025evaluating,vogel2018exploring}. 
In particular, enabling accelerators to operate directly on virtual-memory-managed data introduces three fundamental challenges. First, accelerators require access to physical memory locations while applications operate using virtual addresses. Second, application data is often distributed across non-contiguous physical pages, complicating efficient accelerator data movement. Third, tightly integrated accelerators must maintain memory consistency with processor caches. To address these challenges, we introduce Virtual-Memory-Aware (VMA) method consisting of three complementary mechanisms: a Custom Instruction Virtual Address Translation (CIVAT) for accelerator-side address translation, a Scatter-Gather Page Transfer (SGPT) mechanism for efficient non-contiguous memory access, and a HW-Managed Accelerator Coherence (HMAC) protocol that maintains consistency between accelerator and processor caches. Together, these mechanisms enable accelerators to operate directly on virtual-memory-managed application data without requiring expensive SW-mediated address translation, data copying, or coherence management.

Accelerator execution is also accompanied by invocation and completion overheads \cite{suresh2024mozart}. Unlike CPU execution, which is directly managed by the cores with a simple function call, accelerator execution typically requires traversing multiple SW layers including user-space libraries, system calls, device drivers, and runtime frameworks before computation can begin. Similarly, accelerator completion is commonly communicated through interrupts that trigger additional O/S processing and context switches. While these overheads may appear negligible for long-running accelerators, they can become a significant fraction of total execution time for fine-grained workloads and frequently invoked accelerators. To reduce these costs, we replace expensive Input/Output Control (IOCTL)-based accelerator requests with a lightweight function-call interface that minimizes SW overhead during accelerator invocation. Furthermore, we introduce Interrupt Policy Adaptation (IPA), a runtime mechanism that dynamically selects between polling and interrupt-driven completion handling according to the predicted accelerator execution time. Together, these techniques reduce invocation latency, avoid unnecessary context switches, and lower the O/S tax associated with accelerator execution.

Existing O/S schedulers introduce another source of inefficiency \cite{mack2023cedr,goksoy2021dynamic}. Traditional schedulers were developed for homogeneous multicore systems in which execution resources are largely interchangeable. In heterogeneous systems, however, CPUs and accelerators exhibit widely different performance and energy characteristics. Nevertheless, users are often responsible for explicitly selecting accelerators, while the O/S remains largely unaware of accelerator capabilities and availability. This can lead to suboptimal resource utilization and increased execution time \cite{luk2009qilin,mack2023cedr}. In this work, we demonstrate that treating CPUs and accelerators as a unified pool of computational resources enables more effective scheduling decisions. By dynamically selecting between CPUs and multiple accelerator instances based on workload characteristics and resource availability, the scheduler can improve system utilization and overall application performance. Furthermore, we show that, similar to multicore CPU systems, the benefits of multiple accelerators can only be fully realized when they are managed by an accelerator-aware scheduler.

The key contributions of this paper are summarized as follows:

\begin{itemize}

\item We develop \textsc{\ToolName}, an Field-Programmable Gate Array (FPGA)-based Linux research framework for systematically studying O/S taxes in heterogeneous SoCs, enabling detailed analysis of CPU--accelerator interactions and cross-layer performance bottlenecks.

\item We propose a HW--SW co-designed communication method consisting of CIVAT, SGPT, and HMAC, enabling accelerators to efficiently access virtual-memory-managed application data while achieving up to 260$\times$ lower communication overhead with near-zero HW cost.

\item We reduce accelerator invocation and completion overheads through a lightweight function-call-based execution interface and IPA, a runtime mechanism that dynamically selects between polling and interrupt-driven completion based on predicted accelerator execution time.

\item We extend the O/S scheduler with accelerator-aware resource selection that treats CPUs and multiple accelerator instances as a unified pool of computational resources, achieving up to 2--6$\times$ performance improvement in wall clock time over conventional accelerator-selection approaches.

\item We characterize the system-level behavior of heterogeneous multi-accelerator platforms and demonstrate that application performance depends not only on accelerator speed, but also on O/S overheads, resource availability, and scheduling decisions, highlighting the importance of system-level optimization for future accelerator-rich systems.

\end{itemize}

The remainder of this paper is organized as follows. Section \ref{sec:related} reviews existing frameworks and prior work on O/S taxes. Section \ref{sec:proposed} presents our framework and the techniques employed to reduce these taxes. Section \ref{sec:experimentalresult}  and \ref{sec:result} evaluates the proposed approach, quantifies the resulting performance improvements, and demonstrates how treating accelerators as first-class system resources can substantially reduce the performance gap imposed by O/S overheads.

\section{Related Work and Motivation}
\label{sec:related}

While numerous prior works have addressed individual sources of overhead in heterogeneous systems, including accelerator invocation latency, communication overhead, cache coherence, interrupt handling, and accelerator-aware scheduling, these challenges are typically studied in isolation. Among the closest efforts, Mozart~\cite{suresh2024mozart} demonstrates that SW overheads can substantially reduce the benefits of HW acceleration and introduces the concept of accelerator taxes associated with accelerator communication in NOC and its invocation. Similar to Mozart, we argue that SW overheads play a critical role in determining system performance. However, whereas Mozart primarily focuses on the runtime overheads of supporting disaggregated accelerators, our work takes a broader O/S perspective by systematically characterizing the end-to-end SW overheads encountered by shared-memory accelerators. 


\subsection{Evaluation Framework}

A wide range of infrastructures have been proposed for evaluating heterogeneous systems, each with different trade-offs between flexibility, execution speed, and architectural visibility. Simulation frameworks such as gem5~\cite{binkert2011gem5} and gem5-Aladdin~\cite{shao2016co} provide extensive architectural flexibility but incur low execution speed, limiting full-system Linux evaluation and long-running workloads. FPGA-based platforms, including FireSim~\cite{karandikar2018firesim}, HERO~\cite{kurth2017hero}, ESP~\cite{mantovani2020agile}, LiteX~\cite{kermarrec2020litex}, and commercial FPGA SoCs, provide significantly higher execution speeds but typically rely on large processor cores not optimized for FPGA implementation. This leads to lower achievable frequencies, higher FPGA resource utilization, and limited flexibility for modifying processor architecture and operating-system interactions.

In contrast, our objective is to study operating-system overheads in heterogeneous systems, which requires a highly configurable platform capable of migrating OS functionality and SW "taxes" into HW to improve performance. To achieve this, we build our framework around the lightweight CVA5 soft processor~\cite{matthews2017taiga}. Compared to larger Linux-capable soft processors, CVA5 provides an excellent balance of FPGA resource efficiency, configurability, and performance, leaving sufficient resources for large HW accelerators while remaining fully modifiable~\cite{matthews2018evaluating}. A comparative study of Linux-capable soft processors also found that CVA5 delivers the highest performance among existing designs~\cite{heinz2019catalog}. 

\subsection{Communication Methods}

Traditionally, communication between CPUs and HW accelerators has been performed through a copy-based method which uses SW-managed Direct Memory Access (DMA) transfers using dedicated physically contiguous memory regions shared between the host and the accelerator \cite{meloni2016high,kurth2016mobile,mack2023cedr}. Although this approach requires no additional HW support and performs well for applications with modest data-transfer requirements, SW-managed data copies introduce significant latency and CPU overhead.
To eliminate these SW-managed copies, prior work has explored shared virtual memory (SVM), which allows accelerators to directly access virtual-memory-managed application data \cite{vogel2018exploring,koenig2025evaluating}. SVM has been extensively studied in heterogeneous systems ranging from GPUs and programmable many-core accelerators to FPGA-based computing platforms \cite{amd2017rseries,nvidia2017tx2,vogel2018exploring}. By removing explicit data movement and enabling a shared address space between SW and HW, SVM improves both programmability and performance.



Existing SVM implementations generally rely on IOMMUs, translation caches, and accelerator-side memory-management HW to perform virtual-to-physical address translation \cite{stuecheli2015capi,putnam2014reconfigurable,choi2016quantitative}. While effective, these solutions incur substantial HW complexity and resource overhead, making them difficult to deploy in resource-constrained FPGA-based embedded systems. To address these challenges, several works have proposed lightweight FPGA-oriented SVM solutions. \cite{vogel2018exploring} introduced a configurable FPGA-resident IOMMU and demonstrated up to a 13$\times$ improvement over conventional copy-based communication. More recently, \cite{koenig2025evaluating} evaluated an IOMMU-based shared virtual addressing architecture for RISC-V heterogeneous systems and reported approximately 3$\times$ improvement compared to copy-based method. These results demonstrate the potential of SVM for FPGA accelerators. However, both approaches still rely on dedicated address-translation HW and associated translation overheads. Furthermore, the relatively low effective memory bandwidth reported by recent soft-processor-based IOMMU implementations suggests that lightweight FPGA-oriented SVM architectures have not yet reached their full performance potential.


\subsection{Interrupt/Polling Trade-offs}

Prior work has explored the trade-off between interrupt-driven and polling-based event handling in operating systems, particularly in networking and storage systems. Linux's NAPI framework \cite{salim2001beyond} dynamically switches between interrupts and polling to reduce interrupt overhead under high event rates, while hybrid interrupt-polling schemes have demonstrated that polling can provide higher throughput and lower processing overhead than pure interrupt-driven execution when events occur frequently \cite{salah2007performance,salah2009implementation}. More recently, studies in low-latency storage systems have shown that polling can outperform interrupts in both latency and energy efficiency for fast I/O operations \cite{harris2022poll}. However, these approaches typically make decisions based on device load, traffic characteristics, or fixed thresholds rather than the expected completion time of individual operations.
To the best of our knowledge, prior work has not explored using execution-time prediction to dynamically select between polling and interrupt-driven completion handling for HW accelerators. Our work addresses this gap by leveraging scheduler-provided execution-time estimates to select the completion mechanism that minimizes operating-system overhead for each accelerator invocation.

\begin{figure}[t]
\centering
\includegraphics[width=\columnwidth]{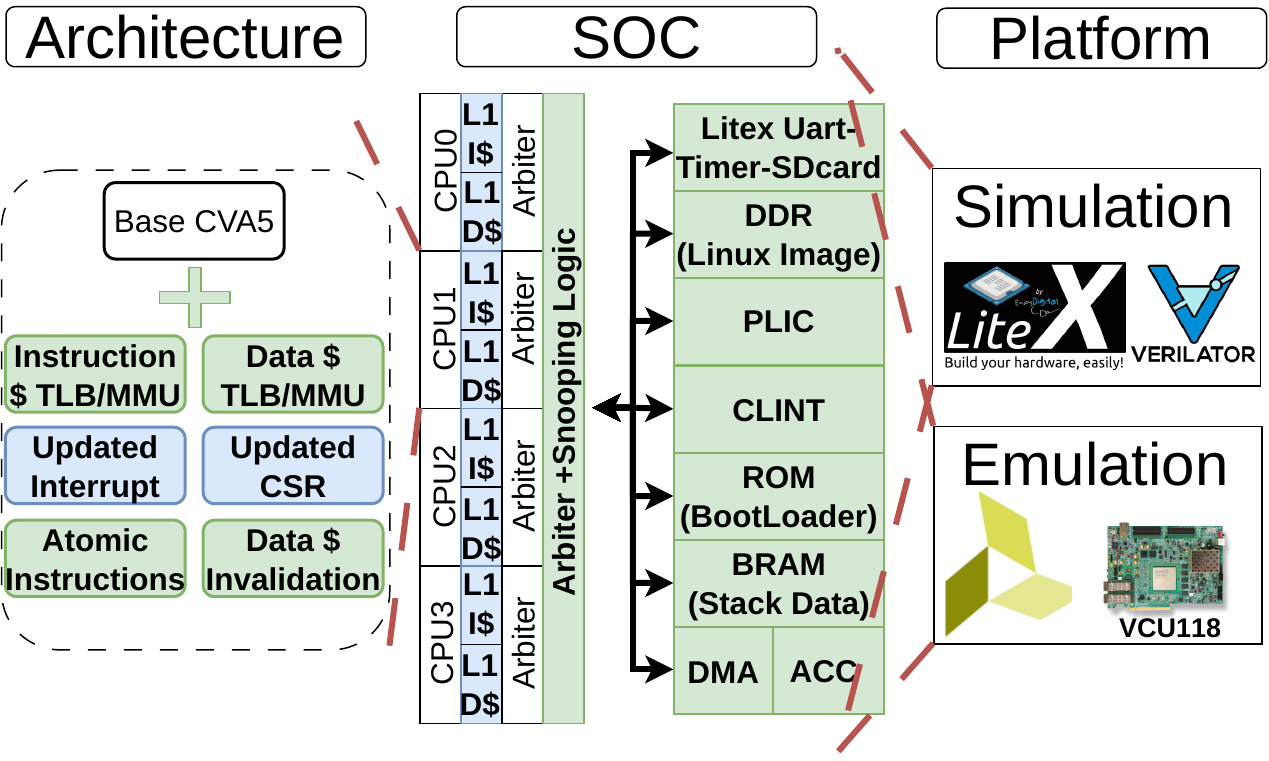}
\caption{Overview of the proposed \ToolName. Green blocks indicate modifications to the original CVA5\cite{matthews2017taiga} architecture and SoC while the blue blocks show the updated blocks.}
\label{fig:framework}
\end{figure}

\subsection{Scheduling}

Scheduling in heterogeneous systems is a well-studied problem. Prior works have demonstrated substantial performance improvements through runtime scheduling across multiple CPUs and GPUs by dynamically or statically selecting the most suitable computing resource for each task \cite{topcuoglu2002performance,
gregg2011dynamic,
choi2013efficient,
kadjo2015control,
augonnet2009starpu,
zhou2014task,
wen2014smart,
bolchini2018runtime,
georgiev2016leo,
aldegheri2018enhancing,
hsieh2019surf}. However, comparatively little attention has been given to the overheads incurred by the scheduling infrastructure itself. More recent frameworks such as IRIS and CEDR \cite{kim2021iris,mack2023cedr} employ worker-based execution models in which the scheduler and each heterogeneous computing resource are managed through dedicated worker threads. While effective, these approaches introduce additional SW overheads that can become significant, particularly for fine-grained accelerator invocations. In this work, we quantify these scheduling overheads and demonstrate techniques to reduce their impact. 

\section{Proposed \ToolName}
\label{sec:proposed}

To systematically study O/S taxes in heterogeneous SoCs, we developed \ToolName, a Linux-capable FPGA-based research framework built on the CVA5 RISC-V processor. CVA5 is a highly configurable RV32IMFD soft processor designed for FPGA deployment and extensibility. To support Linux execution and accelerator-aware O/S research, substantial modifications were made to both the processor architecture and the surrounding SoC infrastructure. These additions include support for virtual memory, privilege modes, interrupt management, cache coherence, multicore execution, and accelerator integration. Figure~\ref{fig:framework} presents an overview of \ToolName, with the modifications to the original CVA5-based system highlighted in green and updated blocks marked in blue.

To enable rapid design-space exploration and system evaluation, \ToolName{} supports both simulation and FPGA emulation. For simulation, we leverage the LiteX platform, which provides fast Verilator-based cycle-accurate simulation of the complete SoC. For FPGA emulation, we implement the SOC using Vivado Block Design and deploy it on the Xilinx VCU118 development board. To maximize SW portability across both environments, the same Universal Asynchronous Receiver-Transmitter (UART), timer, interrupt controller, and peripheral infrastructure are used in both the LiteX and Vivado implementations. Consequently, a single SW stack—including the first-stage bootloader, OpenSBI, device tree, and Linux kernel—can be executed without modification on either platform. The Vivado-based implementation additionally enables the integration of vendor IP blocks, flexible clocking configurations, and high-performance FPGA interfaces, providing a practical environment for evaluating realistic heterogeneous SoC deployments.


\subsection{Communication Tax}
\label{sec:communication}

\begin{figure}[t]
    \centering
    \includegraphics[width=\columnwidth]{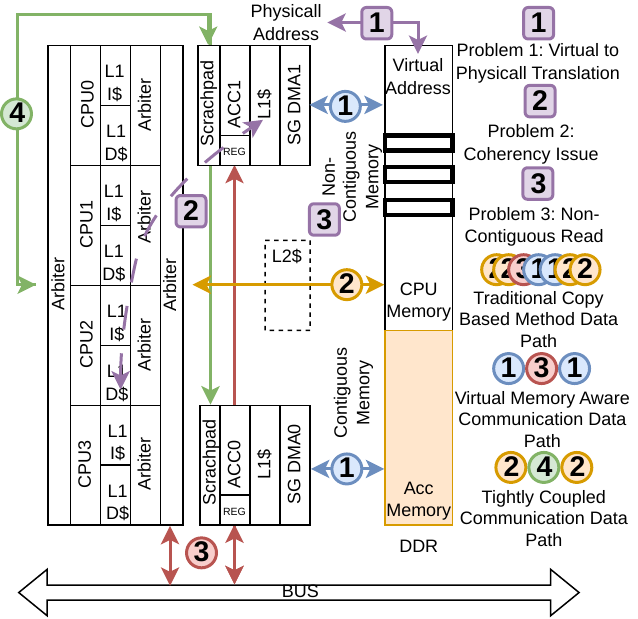}
    \caption{Overview of the three CPU--accelerator communication models supported by the framework and the operating-system challenges associated with each.
    }
    \label{fig:soc}
\end{figure}

The framework supports three communication mechanisms between the CPU and HW accelerators. Regardless of the communication model, three fundamental challenges must be addressed, highlighted by the purple boxes in Figure~\ref{fig:soc}. First, user applications operate on virtual addresses, while accelerators require physical addresses to access memory. Second, data coherence must be maintained between copies of data residing in processor caches and data accessed by accelerators. Finally, O/S memory allocation produces physically non-contiguous pages, requiring a mechanism for accelerators to efficiently access scattered memory regions.

The first communication model (copy-based method) allocates a dedicated memory region for accelerator use. As illustrated by the split DDR memory organization in Figure~\ref{fig:soc}, the CPU and accelerator operate on separate memory spaces. The accelerator region is physically contiguous, non-virtualized, and uncached, allowing direct DMA access without requiring address translation or cache-coherence support. The primary disadvantage of this approach is the additional data movement overhead. Before execution, application data must be copied from the CPU's virtual-memory space into the accelerator-accessible region. After computation completes, the results must be copied back into the application's address space. These extra memory transfers increase execution time and consume memory bandwidth, particularly because accesses to the uncached region incur full DDR latency. Despite these limitations, dedicated accelerator memory remains attractive for storing accelerator-private data structures, such as neural-network weights, lookup tables, and configuration data that are infrequently accessed by the CPU.

The second communication model employs a tightly coupled interface between the CPU and accelerator. In this approach, the accelerator is attached directly to on-chip memory accessible through dedicated uncached load and store operations. CVA5 supports this capability by allowing selected address regions to bypass the cache hierarchy and access FPGA BRAM resources directly. Because all data movement is performed by the CPU, virtual-memory translation, cache coherence, and memory protection are handled transparently by the processor and O/S, eliminating the need for accelerator-side support. The main limitation of this approach is the limited capacity of FPGA on-chip memory. Application data typically resides in DDR memory and must therefore be explicitly transferred between DDR and the accelerator-accessible Block RAM (BRAM). While accesses to the tightly coupled interface can complete in a single cycle, the cost of moving large datasets between DDR and on-chip memory limits scalability. Consequently, this communication model is best suited for fine-grained accelerator invocations and workloads with small working sets.

\subsection{VMA Communication Method}

To eliminate the overheads associated with explicit data copying, the framework also supports a SVM communication model. In this approach, accelerators directly access application data structures using virtual addresses, allowing CPU and accelerator resources to operate on the same memory space without intermediate copies. However, enabling this model requires mechanisms for virtual-to-physical address translation, coherence management, and efficient access to physically non-contiguous memory pages. The design and implementation of this mechanism are described in the following section.

\subsection{Custom Instruction Virtual Address Translation (CIVAT)}

\begin{figure}[t]
    \centering
    \includegraphics[width=\columnwidth]{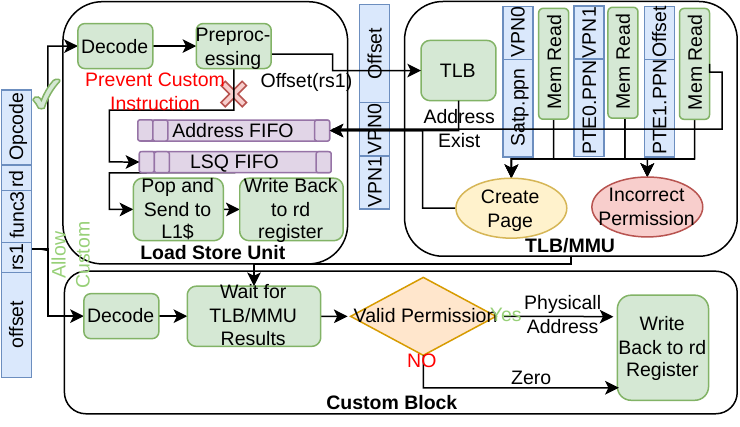}
    \caption{CIVAT instruction datapath. A virtual address supplied in and translated using the processor's existing TLB/MMU mechanisms, and the resulting physical address is forwarded to the custom execution unit for writeback to the destination register.}
    \label{fig:custom_translation}
\end{figure}

\begin{figure}[t]
    \centering
    \includegraphics[width=\columnwidth]{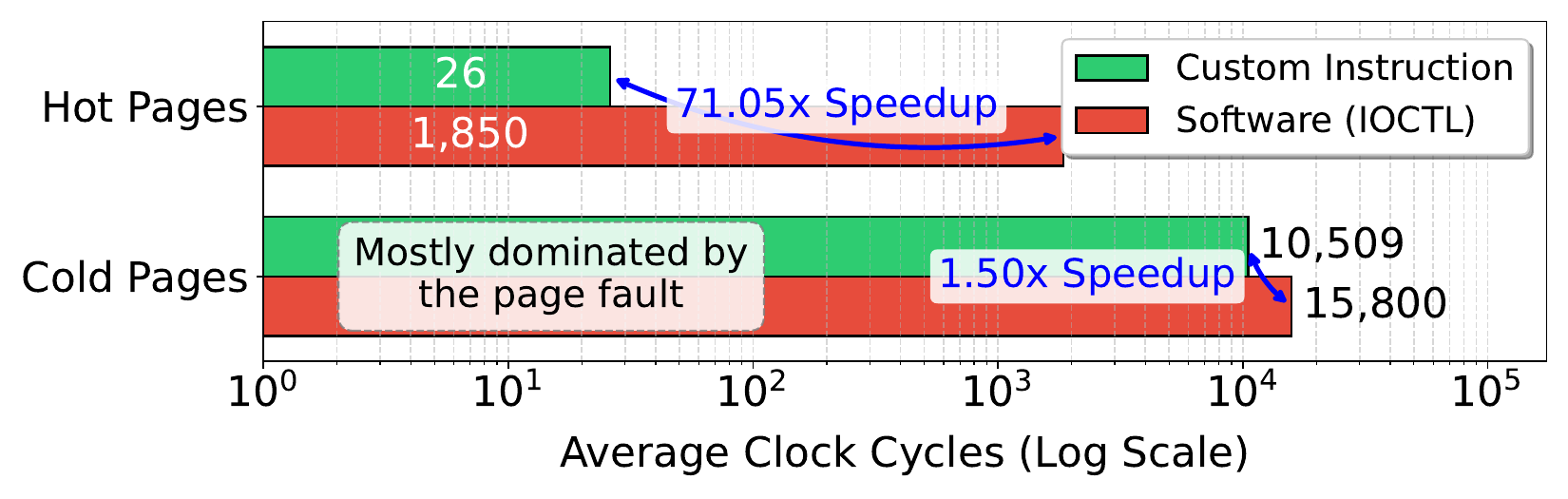}
    \caption{Virtual-to-physical address translation latency using the conventional Linux SW path and the proposed CIVAT instruction. 
    }
    \label{fig:translation_overhead}
\end{figure}

Obtaining the physical address corresponding to a virtual address is a significant source of overhead when accelerators require direct memory access. In Linux, if the target page has not yet been allocated, the processor must first service a page fault and create the corresponding page-table entry. Even when the page is already present, user-space applications cannot directly access physical addresses. Instead, a kernel call is required to walk the page tables, locate the physical page, and return the translation. As shown in Figure~\ref{fig:translation_overhead}, this process requires thousands to tens of thousands of cycles for a single address translation, making it a substantial O/S tax for accelerator communication.

To reduce this overhead, we implemented CIVAT within the CVA5 processor. Figure~\ref{fig:custom_translation} illustrates the required architectural modifications. The instruction uses the custom RISC-V opcode \texttt{0x7B} and accepts a virtual address through the source register (\texttt{rs}). When decoded, the instruction is simultaneously forwarded to the custom execution unit and the load-store unit. The load-store unit performs the normal address translation process using the existing TLB and MMU infrastructure but suppresses the generation of any memory read or write requests. Once translation completes, the resulting physical address is forwarded directly to the custom execution unit, which writes the translated address back to the destination register (\texttt{rd}).

This approach preserves the protection guarantees of the O/S. If the calling process lacks permission to access the requested page, the processor raises a page fault exactly as it would for a conventional memory access. Similarly, if the page has not yet been allocated, the normal page-fault handling mechanism is invoked, allowing Linux to allocate and map the page before the translation is returned. Consequently, the custom instruction exposes only physical addresses that are already accessible to the requesting process while leveraging the existing virtual-memory infrastructure.

Figure~\ref{fig:translation_overhead} shows that the proposed mechanism reduces translation overhead by up to $71\times$ when the page is already present in memory. Even in the page-fault case, where O/S page allocation dominates execution time, the proposed approach still achieves a $1.5\times$ improvement over the conventional SW-based translation path.

\subsection{Scatter-Gather Page Transfer (SGPT)}

To address the challenge of non-contiguous physical pages in virtual-memory systems, we leverage the scatter-gather DMA (SGDMA) engine provided by Vivado and develop a SW driver that automatically converts virtual memory regions into DMA-accessible descriptor chains. Algorithm~\ref{alg:dma_framework_conceptual} presents a simplified view of the driver operation.

The SGDMA engine requires a transmit (TX) chain describing the memory regions to be read and a receive (RX) chain describing the destination regions to be written. Each chain consists of a list of physical-address and transfer-size pairs stored in memory, with the physical address of the chain head provided to the DMA engine. Since the accelerator memory region shown in Figure~\ref{fig:soc} occupies a fixed physical-address range, both the TX and RX descriptor chains are allocated within this region, allowing the DMA HW to access them directly.

To construct these chains, the driver receives a list of arrays or matrices that will either be transferred to the accelerator or populated with accelerator results. As shown in Lines 22--23 of Algorithm~\ref{alg:dma_framework_conceptual}, the driver invokes the \texttt{BuildChain()} procedure to generate the corresponding DMA descriptors. For each matrix, the procedure iterates through all rows and obtains the physical address of the first element using the proposed CIVAT instruction. Because a row may span multiple virtual-memory pages, the driver determines the remaining bytes available in the current 4~KB page and creates a descriptor covering only the contiguous portion of memory within that page. The process then advances to the next page and repeats until the entire row has been described. This procedure converts a virtually contiguous data structure into a sequence of DMA transactions.

An additional advantage of the proposed driver is its support for two-dimensional memory transfers. Rather than requiring SW to manually repack, transpose, or reorient matrix tiles before accelerator execution, the driver can directly describe each row through the descriptor chain. This capability is particularly beneficial for matrix-multiplication accelerators, where input matrices tiles are frequently accessed in different orientations. By allowing the DMA engine to gather data directly from the original matrix layout, the driver eliminates SW-side data rearrangement overhead and reduces unnecessary memory copies.

After the descriptor chains have been generated, the driver programs the SGDMA engine with the addresses of the TX and RX chain heads and initiates the transfer. As a result, the accelerator can efficiently access application data regardless of the underlying physical-page layout, eliminating the requirement for physically contiguous allocations while remaining fully compatible with Linux virtual memory. Unlike traditional approaches that require contiguous DMA buffers or explicit copies into accelerator memory, the proposed SGDMA driver enables accelerators to operate directly on application data allocated by the Linux virtual-memory system.

\begin{algorithm}
\footnotesize
\caption{Scatter-Gather DMA Handling}
\label{alg:dma_framework_conceptual}
\begin{algorithmic}[1]

\Procedure{BuildChain}{$matList, ll$}
    \For{\textbf{each} $mat \in matList$}
        \For{$row \gets 0$ \textbf{to} $rows-1$}
            \State $vaddr \gets \Call{RowStart}{mat,row}$
            \State $bytes \gets \Call{RowSize}{mat}$

            \While{$bytes > 0$}
                \State $paddr \gets \text{VirtToPhys}(vaddr)$
                \State $pageLeft \gets 4096-(paddr \bmod 4096)$
                \State $chunk \gets \min(bytes,pageLeft)$

                \State \Call{Append}{$ll,paddr,chunk$}

                \State $vaddr \gets vaddr + chunk$
                \State $bytes \gets bytes - chunk$
            \EndWhile
        \EndFor
    \EndFor
    \State \Call{FinalizeChain}{$ll$}
\EndProcedure

\Statex

\Procedure{Transfer}{$dst,src,t_{est}$}
    \State \Call{ResetDMA}{}

    \State $tx \gets [src_A,src_B,\dots]$
    \State $rx \gets [dst_A,dst_B,\dots]$

    \State \Call{BuildChain}{$tx,txLL$}
    \State \Call{BuildChain}{$rx,rxLL$}

    \State $ovhd \gets \Call{SetupOverhead}{}$
    \State $irq \gets (t_{est}-ovhd) > T_{switch}$

    \If{$irq$}
        \State $tok \gets \Call{EnableIRQ}{}$
    \EndIf

    \State \Call{SetChainHeads}{$txLL,rxLL$}
    \State \Call{CommitTails}{$txLL,rxLL$}

    \If{$irq$}
        \State \Call{WaitForIRQ}{$tok$}
    \Else
        \While{$\neg\Call{TransferDone}{}$}
        \EndWhile
        \State \Call{ClearFlags}{}
    \EndIf

    \State \Call{StopDMA}{}
\EndProcedure

\end{algorithmic}
\end{algorithm}

\subsection{HW-Managed Accelerator Coherence (HMAC)}


Cache coherence presents another challenge when accelerators access data residing in the CPU's virtual-memory space. CVA5 employs write-through L1 data caches and a snooping-based coherence mechanism to invalidate cache lines modified by other processor cores. A straightforward solution would be to expose every accelerator memory access to the snooping bus so that corresponding cache lines can be invalidated. However, this approach does not scale well as the number of accelerators in the system increases. Furthermore, DMA engines typically issue burst transactions over AXI, meaning that only the starting address and burst length are visible on the bus rather than every individual address accessed by the accelerator. In contrast, processor stores naturally generate individual memory transactions that can be directly observed by the coherence infrastructure.

\begin{figure}[t]
    \centering
    \includegraphics[width=\columnwidth]{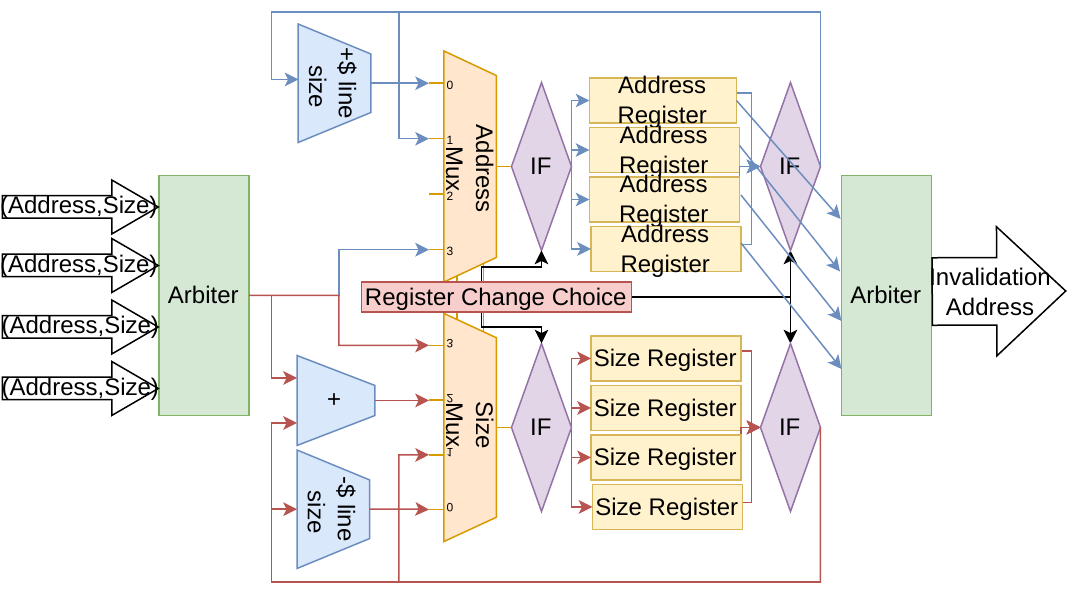}
    \caption{Overview of HMAC, a HW-assisted cache-coherence mechanism that monitors accelerator DMA transactions and generates cache-line invalidation requests for the CVA5 snooping-based cache hierarchy.}
    \label{fig:Coherency}
\end{figure}

To address this challenge, we designed the HW module shown in Figure~\ref{fig:Coherency}. The module monitors AXI transactions generated by each accelerator and extracts the starting address and transfer size of every burst. This information is stored in a dedicated address and size register associated with each accelerator. The coherence unit then generates cache-line invalidation requests on behalf of the DMA engine by traversing the address range covered by the burst. Starting from the burst address, invalidation requests are issued at cache-line granularity until the entire transfer region has been covered. As a result, all cache lines potentially modified by the accelerator are invalidated from the processor caches, ensuring coherence between accelerator-generated data and CPU accesses.

An additional optimization is included to efficiently handle DMA transfers that are split into multiple consecutive AXI bursts. When a newly observed burst begins immediately after the address range of a previously tracked burst, the coherence unit merges the two transactions by extending the stored transfer size rather than creating a new invalidation region. This behavior, illustrated by input two of the size multiplexer in Figure~\ref{fig:Coherency}, makes sure the registers would not override an already existing one.

\subsection{Invocation and Scheduling Tax}

\begin{figure}[t]
    \centering
    \includegraphics[width=\columnwidth]{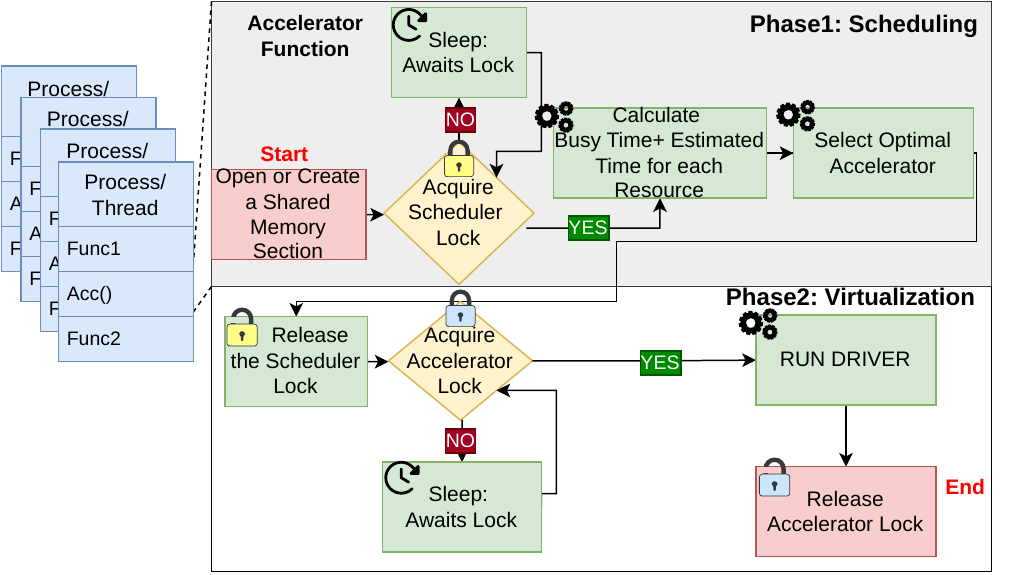}
    \caption{Overview of the 
    accelerator invocation and scheduling path. The design consists of two main components, scheduling and virtualization, coordinated through a shared-memory interface accessible by all processes and threads.}
    \label{fig:Scheduling}
\end{figure}

Previous work has shown that a significant portion of accelerator invocation overhead originates from the SW stack required to access the accelerator, particularly the use of IOCTL calls to communicate with kernel-space drivers. To reduce this invocation tax, we designed the accelerator interface as a user-space library rather than a traditional device driver. As illustrated in Figure~\ref{fig:Scheduling}, applications invoke accelerators through a function call, eliminating the kernel transition and IOCTL overhead associated with conventional accelerator frameworks.

We also investigated the scheduling mechanisms employed by existing open-source heterogeneous computing frameworks and observed that many dedicate a separate scheduling process or worker thread in addition to one or more worker threads responsible for accelerator management. While this design simplifies implementation, it introduces additional O/S overheads. In many workloads, scheduling decisions occur infrequently, causing the scheduler thread to spend most of its execution time waiting for work while still consuming system resources through polling, context switches, or synchronization operations.

To eliminate this scheduler tax, we integrated the scheduling logic directly into the accelerator invocation path, as shown in Figure~\ref{fig:Scheduling}. Instead of forwarding requests to a dedicated scheduler process, the application thread invokes the scheduler immediately before accelerator execution. This approach removes the need for a separate scheduler thread, reducing context-switch overhead and eliminating idle scheduler execution.

Integrating scheduling into the application thread introduces concurrency challenges because scheduling decisions rely on shared system-wide state, including accelerator availability and performance statistics. To ensure correctness, we maintain scheduler and accelerator locks within a shared data structure accessible by all processes. These locks serialize access to shared scheduling information while allowing competing processes to sleep rather than spin while waiting for accelerator resources. As a result, the framework avoids both the overhead of dedicated scheduler threads and the CPU utilization penalties associated with busy-wait synchronization.

\subsection{Interrupt Tax}

Another source of O/S overhead is the context switch associated with accelerator completion interrupts. In a conventional interrupt-driven execution model, every accelerator completion generates an interrupt that forces the O/S to perform interrupt handling, sleep the current process, and execute one or more context switches. For short accelerator executions, this overhead can become comparable to or even exceed the accelerator execution time itself, causing the processor to spend a significant fraction of its time performing O/S work rather than useful computation.

Our experiments revealed that no single completion strategy is optimal across all workloads. For short-running accelerator tasks, actively polling for completion is often more efficient than sleeping and waiting for an interrupt because it avoids the context-switch overhead entirely. In contrast, for long-running accelerator executions, interrupt-based completion is preferable because it allows the processor to execute other work while the accelerator is running. Consequently, we implemented IPA that dynamically selects between polling and interrupt-driven execution.

The proposed policy leverages the execution-time estimates already generated by the scheduler. As shown in Lines 31--37 of Algorithm~\ref{alg:dma_framework_conceptual}, the estimated accelerator execution time is compared against a predefined threshold. If the expected execution time exceeds this threshold, the process is suspended and later resumed through an interrupt. Otherwise, the application polls the accelerator completion status directly, avoiding the overhead of interrupt processing and context switching.

A practical challenge is that accelerator interrupts are connected to the PLIC and managed through a Linux device driver. Consequently, an interrupt would still trigger O/S handling even when polling is selected. To avoid this unnecessary overhead, the framework dynamically enables or disables accelerator interrupts before execution, as shown in Lines 25--28 of Algorithm~\ref{alg:dma_framework_conceptual}. When polling is selected, the interrupt source is disabled, preventing the generation of an unnecessary context switch. When interrupt-driven execution is selected, the interrupt is enabled and normal O/S notification is used. This approach ensures that the completion mechanism itself does not introduce avoidable O/S overhead.

\section{Experimental Setup}
\label{sec:experimentalresult}

We evaluate the proposed framework on a Xilinx VCU118 FPGA platform with all system components operating at 100~MHz. The system boots using the LiteX built-in BIOS as the first-stage bootloader, followed by OpenSBI v0.8 and Linux kernel v5.18.0-rc7. All experimental results were collected using this configuration.

\subsection{Framework}

The baseline evaluation builds upon the CVA5 processor core, configured with two-way set-associative instruction and data caches (512 lines, 8-word width), a 512-entry branch predictor, an 8-entry return address stack, and 64-entry TLBs. To support Linux execution, some modifications and updates were made to the original CVA5 architecture. As detailed in Table~\ref{tab:resource_usage}, extending the baseline CVA5 to a full Linux-capable core increases resource utilization by only 1 BRAM and 211 CLBs (805 versus 594 CLBs), with no additional DSP footprint. The surrounding framework infrastructure—including the multicore wrapper, PLIC, CLINT, and LiteX IP—consumes only 615 CLBs total without using BRAMs or DSPs.

This lightweight hardware modification enables efficient OS-level execution without incurring significant architectural latency. As shown in Table~\ref{tab:platform_baselines}, low-level system operations inside the framework remain minimal: a null system call (\texttt{lat\_syscall}) resolves in 297 clock cycles ($2.97\,\mu\text{s}$). Furthermore, memory throughput benchmarks (\texttt{bw\_mem}) demonstrate low-overhead data handling, requiring 0.37 cycles/byte for write operations and 1.50 cycles/byte for reads.

Building on this optimized single-core baseline, the framework scales predictably in multicore configurations. Table~\ref{tab:cva5_coremark_scaling} presents the parallel CoreMark execution results across up to four cores. Crucially, the software and framework layer introduce negligible overhead: individual core performance remains pinned at $2.59\text{ CoreMark/MHz}$, matching the reported baremetal baseline\cite{heinz2019catalog}. Doubling the workload from one to two cores increases total execution time by only 0.18\,s (1.1\%), while a four-core configuration maintains 94.7\% scaling efficiency, confirming the overall framework scalability.

\begin{table}[htbp]
\centering
\caption{CVA5 multi-Core scaling and CoreMark performance at 100 MHz}
\label{tab:cva5_coremark_scaling}
\resizebox{\columnwidth}{!}{%
    \begin{tabular}{|l|c|c|c|c|c|c|}
    \hline
    \textbf{Core} & \textbf{Parallel} & \textbf{Total} & \textbf{Elapsed} & \textbf{CoreMark} & \textbf{CoreMark/MHz} & \textbf{Scaling} \\
    \textbf{Config.} & \textbf{Threads} & \textbf{Iters} & \textbf{Time (s)} & \textbf{Score} & \textbf{(at 100 MHz)} & \textbf{Eff. (\%)} \\
    \hline
    1 Core  & 1 & 4,000  & 15.45 & 258.88 & 2.59 & 100.0\% \\
    \hline
    2 Cores & 2 & 8,000  & 15.63 & 511.71 & 5.12 & 98.8\% \\
    \hline
    3 Cores & 3 & 12,000 & 15.78 & 760.55 & 7.61 & 97.9\% \\
    \hline
    4 Cores & 4 & 16,000 & 16.32 & 980.69 & 9.81 & 94.7\% \\
    \hline
    \end{tabular}%
}
\end{table}

\begin{table}[htbp]
\centering
\caption{Framework overhead and memory throughput at 100 MHz}
\label{tab:platform_baselines}
\begin{tabular}{|l|c|c|}
\hline
\textbf{Operation /} & \textbf{Raw} & \textbf{Performance in} \\
\textbf{Benchmark} & \textbf{Metric} & \textbf{Clock Cycles} \\
\hline
\textbf{\texttt{lat\_syscall null}} & $2.97\ \mu\text{s}$ & 297 cycles \\
\hline
\textbf{\texttt{bw\_mem 1m wr}} & $271.09\text{ MB/s}$ & 0.37 cycles/byte \\
\hline
\textbf{\texttt{bw\_mem 1m rd}} & $66.59\text{ MB/s}$ & 1.50 cycles/byte \\
\hline
\textbf{\texttt{bw\_mem 1m rdwr}} & $53.76\text{ MB/s}$ & 1.86 cycles/byte \\
\hline
\end{tabular}
\end{table}


\begin{table}[htbp]
\centering
\caption{HW resource utilization breakdown}
\label{tab:resource_usage}
\begin{tabular}{|l|c|c|c|c|c|}
\hline
\textbf{Component} & \textbf{DSP} & \textbf{BRAM} & \textbf{CLB} & \textbf{LUT} & \textbf{FF} \\
\hline
\rowcolor{gray!15} \textbf{Baseline CVA5 (1 Core)} & 4 & 19.5 & 594 & 2905 & 1908 \\
\hline
\rowcolor{gray!15} \textbf{Linux CVA5 (1 Core)}    & 4 & 20.5 & 805 & 4593 & 2745 \\
\quad \textit{CIVAT}                               & -- & --  & 26  & 2    & 38   \\
\hline
\rowcolor{gray!15} \textbf{Multicore Wrapper}    & 0 & 0    & 294 & 1001   & 448   \\
\quad \textit{HMAC}                                & -- & --  & 138 & 422  & 126  \\
\hline
\rowcolor{gray!15} \textbf{PLIC}                   & 0 & 0    & 101 & 489  & 223  \\
\hline
\rowcolor{gray!15} \textbf{CLINT}                  & 0 & 0    & 88  & 324  & 365  \\
\hline
\rowcolor{gray!15} \textbf{Litex IP}               & 0 & 0    & 132 & 587  & 607  \\
\hline
\end{tabular}
\end{table}

\subsection{Applications}

To evaluate the solution to the proposed taxes, we implemented four HW accelerators and four representative applications generating accelerator requests. The benchmark suite includes two communication-bound and two compute-bound applications, enabling evaluation of invocation mechanism, interrupt policy, and scheduler. Table~\ref{tab:benchmarks} summarizes the characteristics of a single application execution; additional repetitions used in specific experiments are stated separately.

\begin{table}[t]
\centering
\caption{Benchmark characteristics. Calls denotes the number of accelerator invocations, Avg. Tx the average transferred data per invocation, CPU Time the CPU-only execution time, ACC Total the accelerator-based execution time including communication overhead, HW the accelerator execution time, and $\times$ the speedup over CPU execution. The unit of time in this table is in seconds.}
\label{tab:benchmarks}
\setlength{\tabcolsep}{2.2pt}
\begin{tabular}{|l|r|r|c|r|r|r|r|}
\hline
App & Calls & Avg. Tx & HW & CPU Time & ACC Total & HW & $\times$ \\
\hline
MM & 40 & 49KB & GEMM64 & 4004 & 4003 & 16.8 & 238 \\
\hline
MM & 40 & 196KB & GEMM128 & 4004 & 4003 & 5.2 & 770 \\
\hline
MC & 10KB & 125K & FIFO & 40.8 & 40.4 & 4.5 & 9.0 \\
\hline
PD & 512 & 1KB & FFT & 4.6 & 3.0 & 0.2 & 15 \\
\hline
\end{tabular}
\end{table}

The communication-bound applications are:
\begin{itemize}
    \item \textbf{Pulse Doppler(PD):} Performs FFT and inverse FFT operations using the Xilinx Vivado FFT IP. Since the FFT core is fully pipelined, performance is primarily limited by data movement rather than computation. As shown in Table~\ref{tab:benchmarks}, each invocation transfers approximately 1~KB and achieves a 15$\times$ speedup over CPU execution.

    \item \textbf{Memory Copy(MC):} Uses a FIFO connected directly to the DMA engine to isolate communication overhead without performing computation. Random-sized transfers between 1~B and 250~KB (125~KB average) achieve a 9$\times$ speedup over SW \texttt{memcpy}. For maximum-size transfers of 1~MB, the DMA-based implementation achieves a 26$\times$ speedup compared to CPU \texttt{memcpy}.
\end{itemize}

The compute-bound applications are:
\begin{itemize}
    \item \textbf{Matrix Multiplication(MM):} Implements a tiled GEMM accelerator in Vivado HLS with AXI-Stream interfaces connected to the DMA engine. The benchmark performs $512\times512$ matrix multiplications while varying the tile size to evaluate different computation-to-communication ratios. Tile sizes of 64 and 128 achieve 238$\times$ and 770$\times$ speedups, respectively, over the CPU implementation.

    \item \textbf{Synthetic Accelerator(SA):} A configurable accelerator with programmable execution latency. It emulates accelerators with different computation times and speedups, enabling controlled evaluation of the scheduler and the proposed Interrupt Policy Adaptation (IPA). Each invocation uses a random latency, allowing requests to favor either polling or interrupt-based completion.
\end{itemize}


\subsection{Scheduling Policies}

We tested five scheduling policies to find the most effective one.

    
    

    

\begin{itemize}
    \item \textbf{CPU Only(CO):} Every task is executed on available CPUs, regardless of accelerator availability or system load.

    \item \textbf{Accelerator Only(AO):} Every task is dispatched to one designated accelerator. If the accelerator is busy, incoming tasks wait until it becomes available.

    \item \textbf{Busy-Aware (Fuse):} Adapted from Fuse~\cite{ismail2011fuse}, this policy dispatches a task to the accelerator whenever it is idle; otherwise, the task is immediately executed on the CPU.

    \item \textbf{Random:} For each task, the scheduler randomly selects one of the available computing resources (CPU or accelerator) without considering execution time, resource utilization, or queue occupancy.

    \item \textbf{Earliest Finish Time (EFT):} Our proposed scheduler estimates the completion time on every available computing resource and assigns the task to the resource predicted to finish first. The estimate accounts for both the expected execution time of the task and the current workload of each resource, enabling dynamic load balancing across heterogeneous processors and accelerators.
\end{itemize}

\section{Experimental Results}
\label{sec:result}

\subsection{Communication tax results}

\begin{figure}[t]
    \centering
    \includegraphics[width=\columnwidth]{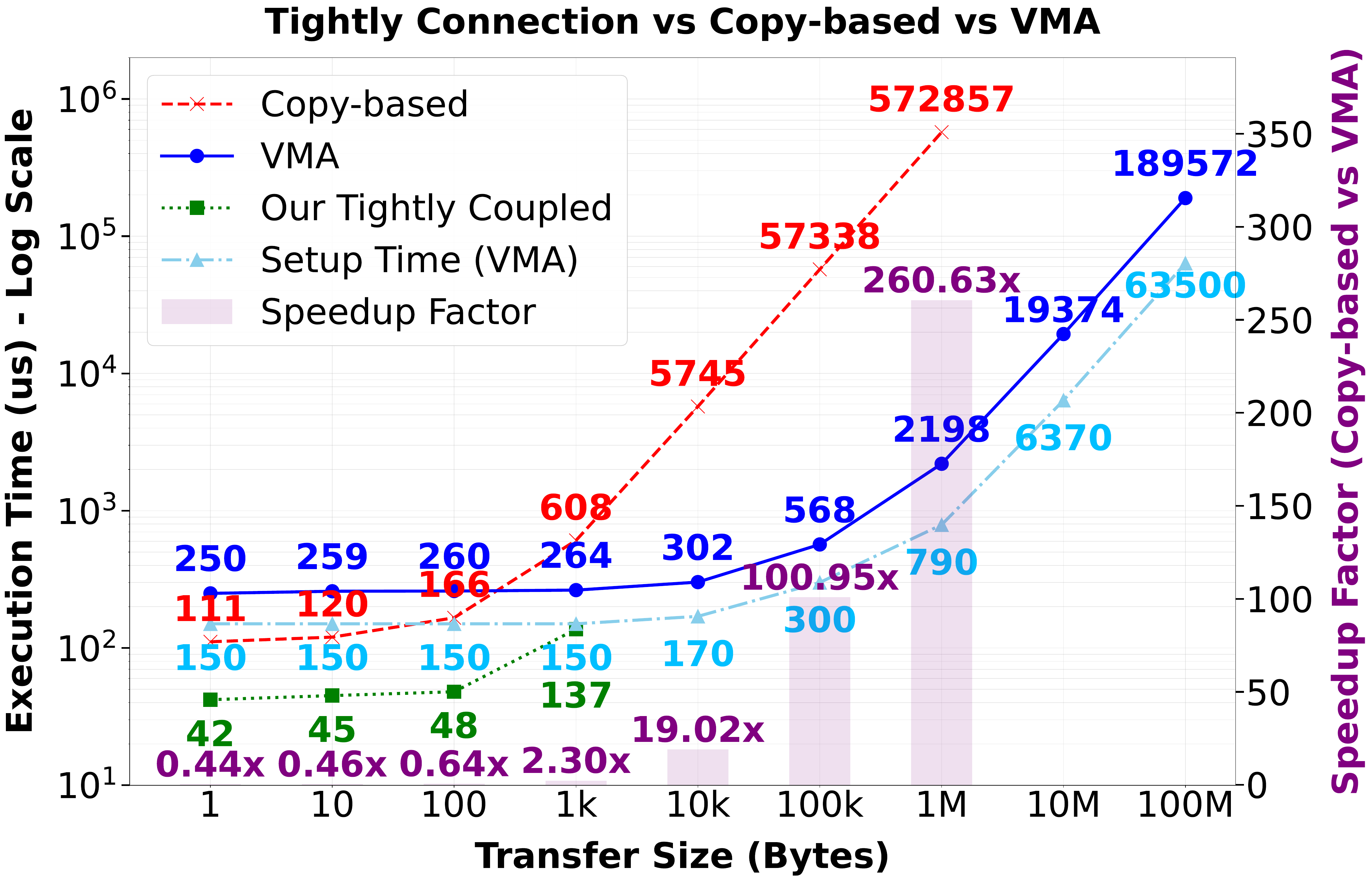}
    \caption{Communication latency of the three accelerator communication mechanisms supported by \ToolName. 
    }
    \label{fig:Comunication_Results}
\end{figure}

\begin{table}[t]
\centering
\caption{Performance comparison of copy-based, the previous IOMMU-based SVM implementation~\cite{koenig2025evaluating}, and the VMA method. 
}
\label{tab:comm_comparison}
\scriptsize
\renewcommand{\arraystretch}{1.1}
\setlength{\tabcolsep}{3pt}
\begin{tabular}{|l|c|c|c|c|}
\hline
\textbf{Metric} &
\textbf{CPU} &
\textbf{SVM} &
\textbf{Our} &
\textbf{Max} \\
&
\textbf{Copy-based} &
\textbf{\cite{koenig2025evaluating}} &
\textbf{VMA} &
\textbf{Performance} \\
\hline

Data Size &
1 MB &
64 KB &
\textbf{1 MB} &
1 MB \\
\hline

Bus Width &
32 &
32 &
\textbf{64} &
64 \\
\hline

Cycles &
57.29M &
2.00M &
\textbf{219.8K} &
140.8K \\
\hline

Cycles/B &
57.29 &
30.52 &
\textbf{0.220} &
0.141 \\
\hline

BW (MB/s) &
1.75 &
3.28 &
\textbf{454.96} &
710.23 \\
\hline

Peak BW &
400 &
400 &
\textbf{800} &
800 \\
\hline

Eff. (\%) &
0.44 &
0.82 &
\textbf{56.87} &
88.78 \\
\hline

Speedup &
1.00$\times$ &
1.88$\times$ &
\textbf{260.63$\times$} &
406.86$\times$ \\
\hline

\end{tabular}
\end{table}

We evaluated the three communication mechanisms introduced in Section~\ref{sec:communication}. Figure~\ref{fig:Comunication_Results} compares their communication latency across different transfer sizes. The tightly coupled interface provides the lowest latency for small transfers because the accelerator communicates directly with on-chip BRAM without accessing external memory. However, its scalability is fundamentally limited by the available on-chip memory resources. In our implementation, the maximum transferable data size is only 1~KB.

The copy-based communication method exhibits significantly higher latency, and the performance gap widens as the transfer size increases. This approach requires the CPU to explicitly copy data between the application's virtual memory and the accelerator's dedicated memory region, causing the communication overhead to grow linearly with the amount of transferred data. Furthermore, the maximum transferable data size is limited by the size of the reserved accelerator memory in DDR.

In contrast, the proposed VMA method combines the low SW overhead of direct memory sharing with the scalability of off-chip memory. Since the accelerator directly accesses the application's virtual memory, no explicit memory copies are required, and the transfer size is no longer constrained by a dedicated accelerator memory region. As shown in Figure~\ref{fig:Comunication_Results}, VMA outperforms both previous communication methods for large transfers while remaining scalable to arbitrarily large data sizes.

Figure~\ref{fig:Comunication_Results} also separates the communication time into address translation and data transfer. Interestingly, approximately 33\% of the total communication time is spent translating virtual addresses, while the remaining 66\% is devoted to the actual DMA transfers. This observation 
suggest that further optimization of the translation mechanism could substantially improve performance.

Table~\ref{tab:comm_comparison} compares our implementation against both the conventional CPU copy-based approach and the only previously published SVM implementation for RISC-V soft processors~\cite{koenig2025evaluating}. VMA achieves an effective bandwidth of 454.96~MB/s, corresponding to 56.9\% of the theoretical 64-bit memory bandwidth. Without address translation, the same communication path reaches 710.23~MB/s (88.8\% efficiency).

Compared to the conventional copy-based approach, the proposed VMA method reduces the communication overhead by up to $260\times$. It also substantially outperforms the previous IOMMU-based SVM implementation~\cite{koenig2025evaluating}, requiring only 0.22 cycles/byte compared to 30.52 cycles/byte reported in their work while the CIVAT consumes only 26 CLBs and HMAC 138 \ref{tab:resource_usage}. The results for the previous SVM implementation are taken directly from the published paper, as we did not reimplement their design.


\subsection{Schedluing Tax}

\begin{figure}[t]
    \centering
    \includegraphics[width=\columnwidth]{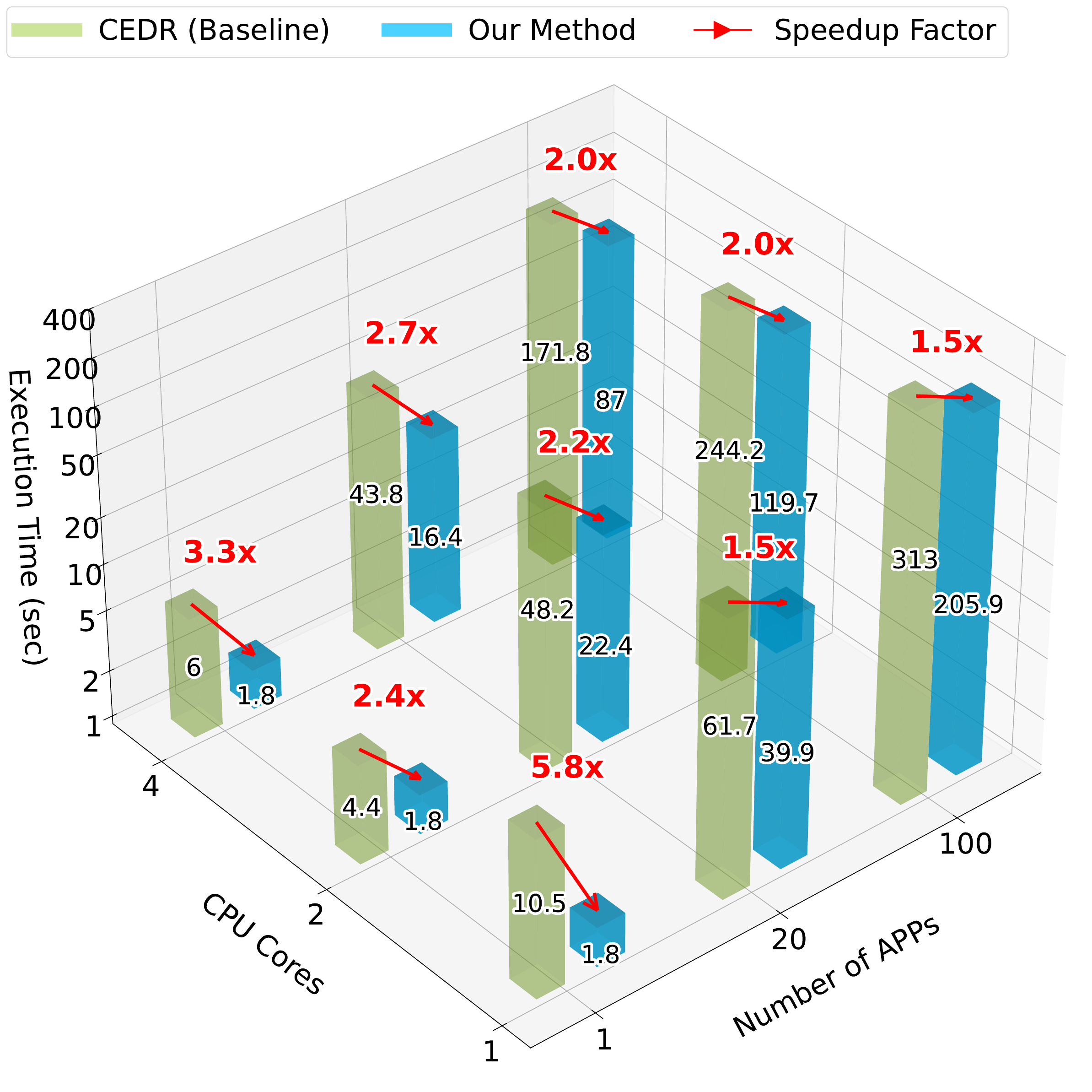}
    \caption{Performance comparison between the proposed scheduler and the worker-based scheduler used in CEDR~\cite{mack2023cedr}. 
    }
    \label{fig:SchedulingComparision}
\end{figure}

Figure~\ref{fig:SchedulingComparision} compares the performance of our scheduler with the worker-based scheduler used by CEDR~\cite{mack2023cedr} using the PD benchmark. Across all configurations, our scheduler improves application wall clock time by $1.5\times$--$5.8\times$.

The largest improvement is observed when the system executes only a small number of applications. For example, with a single CPU core and a single application, our framework achieves a $5.8\times$ speedup. In this scenario, there is little opportunity for parallel execution, making the dedicated scheduler thread in CEDR largely idle while still consuming CPU resources. Furthermore, CEDR allocates an additional worker thread for each HW accelerator, which also remains mostly unused. As a result, a significant fraction of the available processing resources is devoted to scheduler management rather than useful application execution.

As the number of concurrent applications increases, the scheduler and accelerator worker threads become increasingly utilized, reducing the relative overhead of the worker-based design. Consequently, the performance gap narrows. Nevertheless, even when executing 100 PD applications simultaneously, our scheduler still achieves a $1.5\times$--$2\times$ improvement over CEDR by eliminating the scheduler worker thread and reducing scheduling overhead.

Finally, PD is primarily CPU-bound because the FFT accelerator is fully pipelined and spends relatively little time computing compared to the SW processing surrounding each invocation. Consequently, increasing the number of CPU cores consistently improves throughput for both scheduling approaches.

\subsection{Communication Bound Applicaions}

\begin{figure}[t]
    \centering
    \includegraphics[width=\columnwidth]{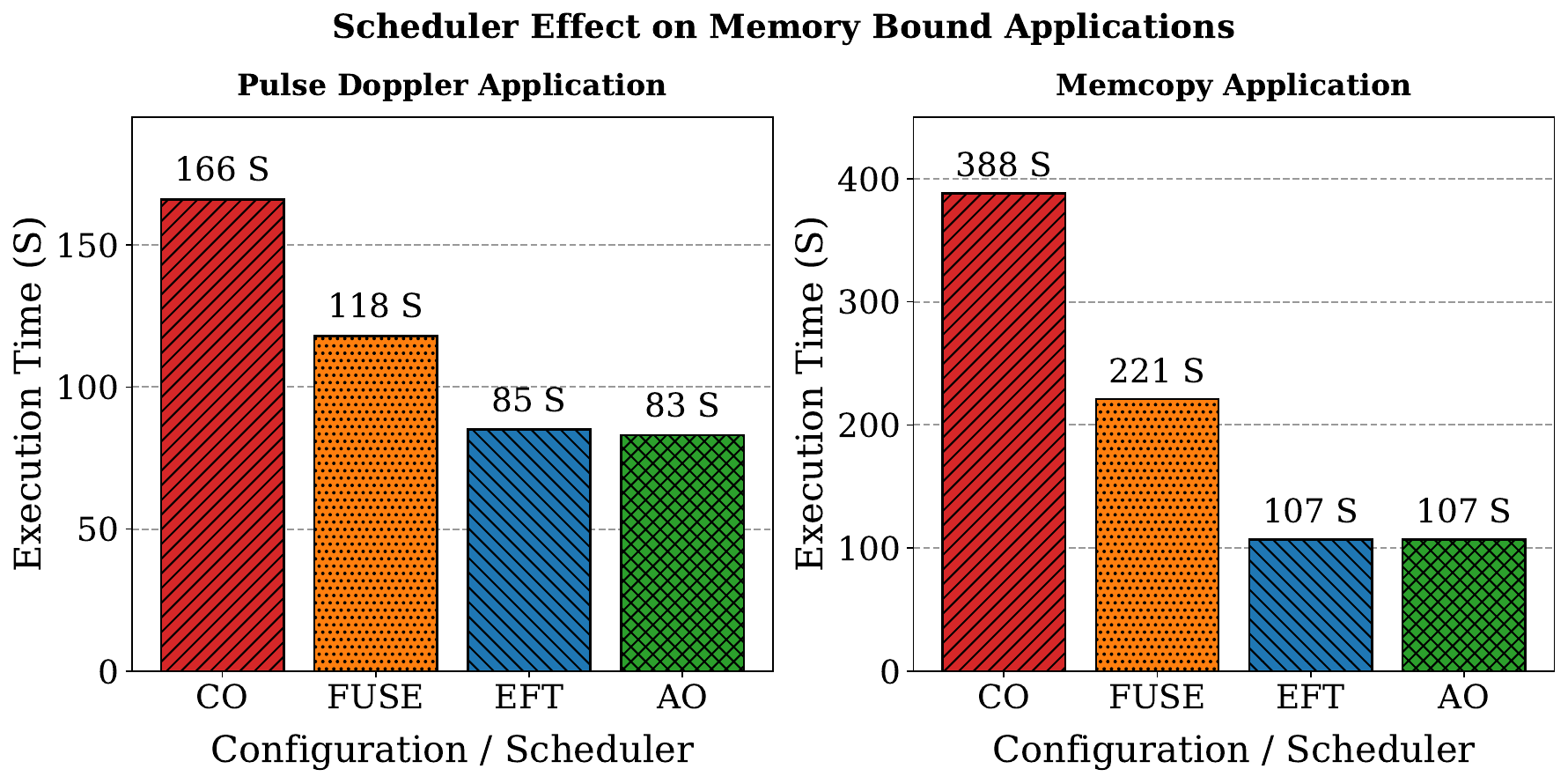}
    \caption{Different Scheduler effect on the overall application execution time. For the PD there are 100 of calls while for the MC there is 20.}
    \label{fig:Scheduler_Memory_Bound}
\end{figure}

Figure \ref{fig:Scheduler_Memory_Bound} presents the results for two communication-bound applications. For the PD application, the SoC was configured with two FFT accelerators available to the schedulers; similarly, the Memcpy application was configured with two FIFO accelerators. However, in both cases, advanced scheduling yields no positive impact on performance. Because bus bandwidth is constrained, data transfers from one accelerator force the other to wait. As a result, throughput remains virtually identical to a system with AO policy.

\subsection{Computation Bound Applicaions}

\begin{figure}[t]
    \centering
    \includegraphics[width=\columnwidth]{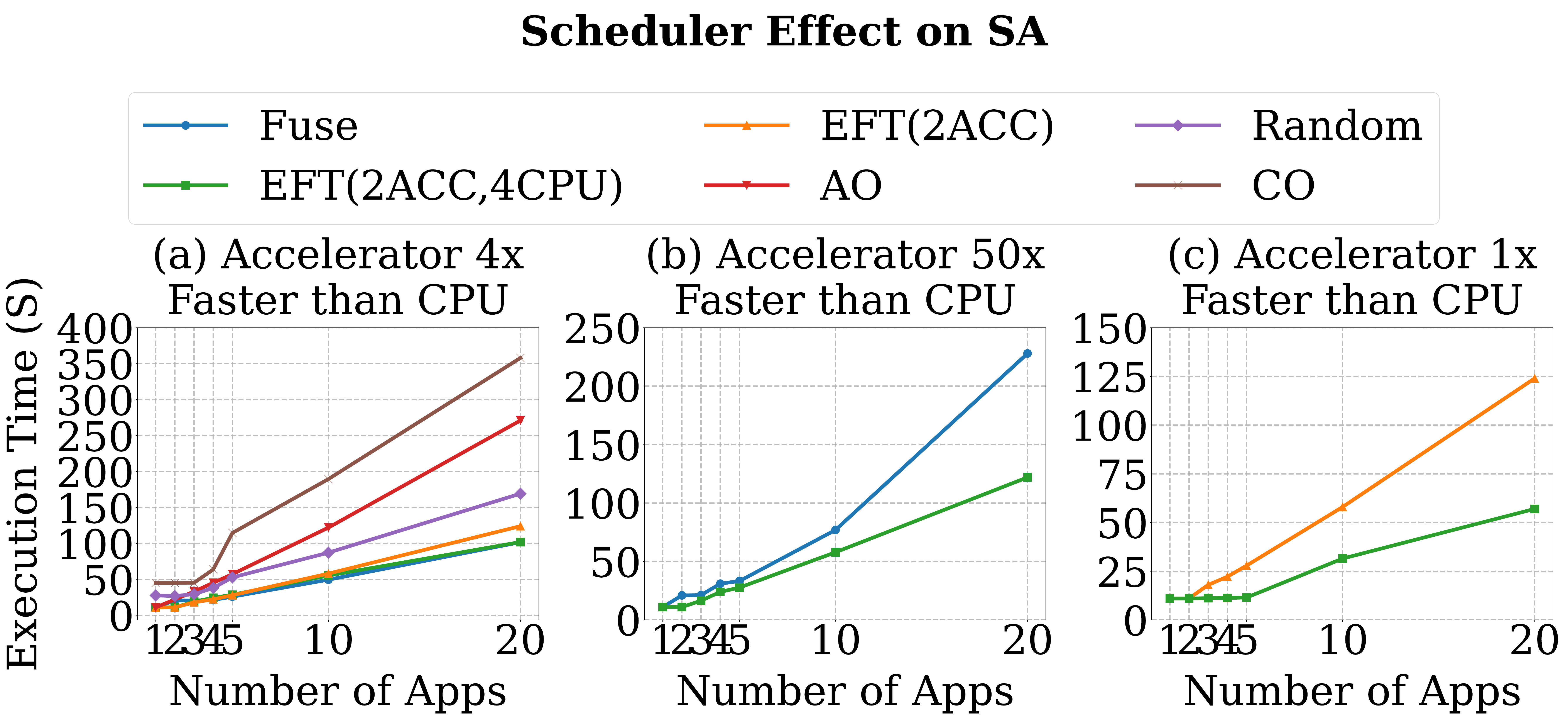}
    \caption{Performance of different scheduling policies for SA as the accelerator speedup relative to the CPU varies. 
    }
    \label{fig:Compute_Bound_Custom}
\end{figure}

For compute-bound workloads, we evaluate both MM and SA. Figure~\ref{fig:Compute_Bound_Custom} illustrates the performance of the evaluated scheduling policies under different accelerator speedups relative to the CPU.

Across all configurations, the proposed EFT scheduler achieves peak performance by dynamically scheduling across all available computing resources. For example, when an accelerator is $4\times$ faster than the CPU, fully configured EFT significantly outperforms both the AO policy and EFT(2ACC)—which restricts decisions to two accelerators. This demonstrates that for compute-bound workloads, co-optimizing CPUs and accelerators yields substantially higher throughput than relying on accelerators alone.

An interesting observation is the strong performance of the Fuse scheduler despite its simple heuristic. When the accelerator provides a moderate speedup, immediately executing tasks on the CPU whenever the accelerator is busy is often sufficient to achieve performance close to that of EFT. However, this heuristic becomes ineffective as the accelerator speedup compare to the CPU increases. In these cases, executing a task on the CPU may take significantly longer than waiting for the accelerator to become available, making Fuse's scheduling decision suboptimal.

Conversely, when the CPU and accelerator provide similar execution times, exploiting both resources becomes increasingly beneficial. In this regime, scheduling tasks across both CPUs and accelerators nearly doubles the achievable throughput compared to policies that dispatch work exclusively to accelerators, demonstrating the importance of using all the available resources in the system.

\begin{figure}[t]
    \centering
    \includegraphics[width=\columnwidth]{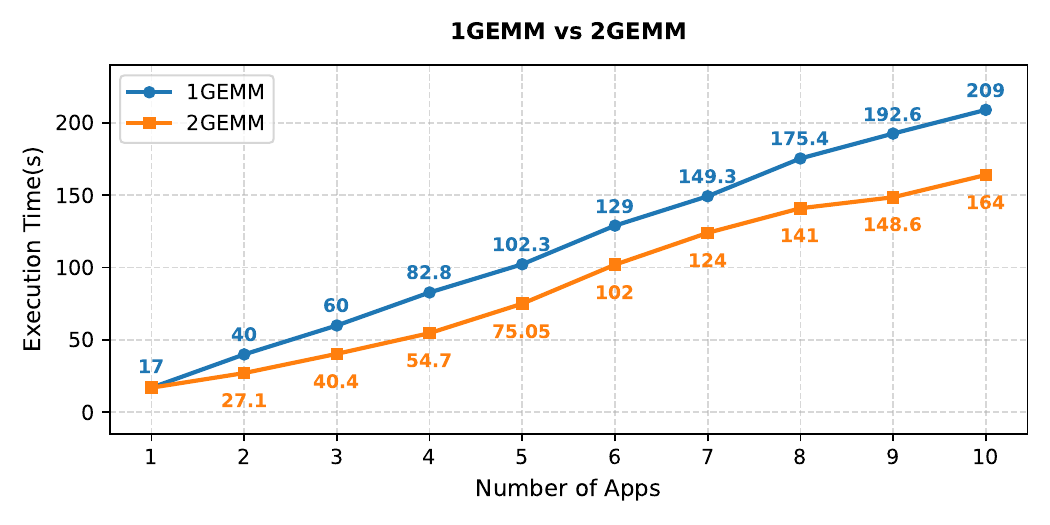}
    \caption{Effect of accelerator replication on a tiled MM workload. 
    }
    \label{fig:GEMM}
\end{figure}

To evaluate the scheduler on a realistic workload, we implemented MM application.
Figure~\ref{fig:GEMM} evaluates the impact of scheduling across multiple GEMM64 accelerators. We compare systems containing one and two identical GEMMs with the tile size of 64 using the proposed EFT scheduler. When multiple applications execute concurrently, distributing requests across two accelerators improves overall throughput by approximately $21\%$ compared to a single-accelerator configuration. These results demonstrate that accelerator replication provides an additional dimension of parallelism. Similar to increasing the number of processor cores, adding multiple accelerators allows independent application requests to execute concurrently, enabling the scheduler to further increase system throughput.

\subsection{Interrupt+Polling}


Our experiments showed that polling consistently outperforms interrupts for short-running workloads such as FFT and memory-copy operations, where the context-switch overhead is comparable to the accelerator execution time. In contrast, interrupt-driven execution is preferable for compute-bound applications, where accelerator execution dominates the overall latency and the processor can perform useful work while waiting for completion.

\begin{figure}[t]
    \centering
    \includegraphics[width=\columnwidth]{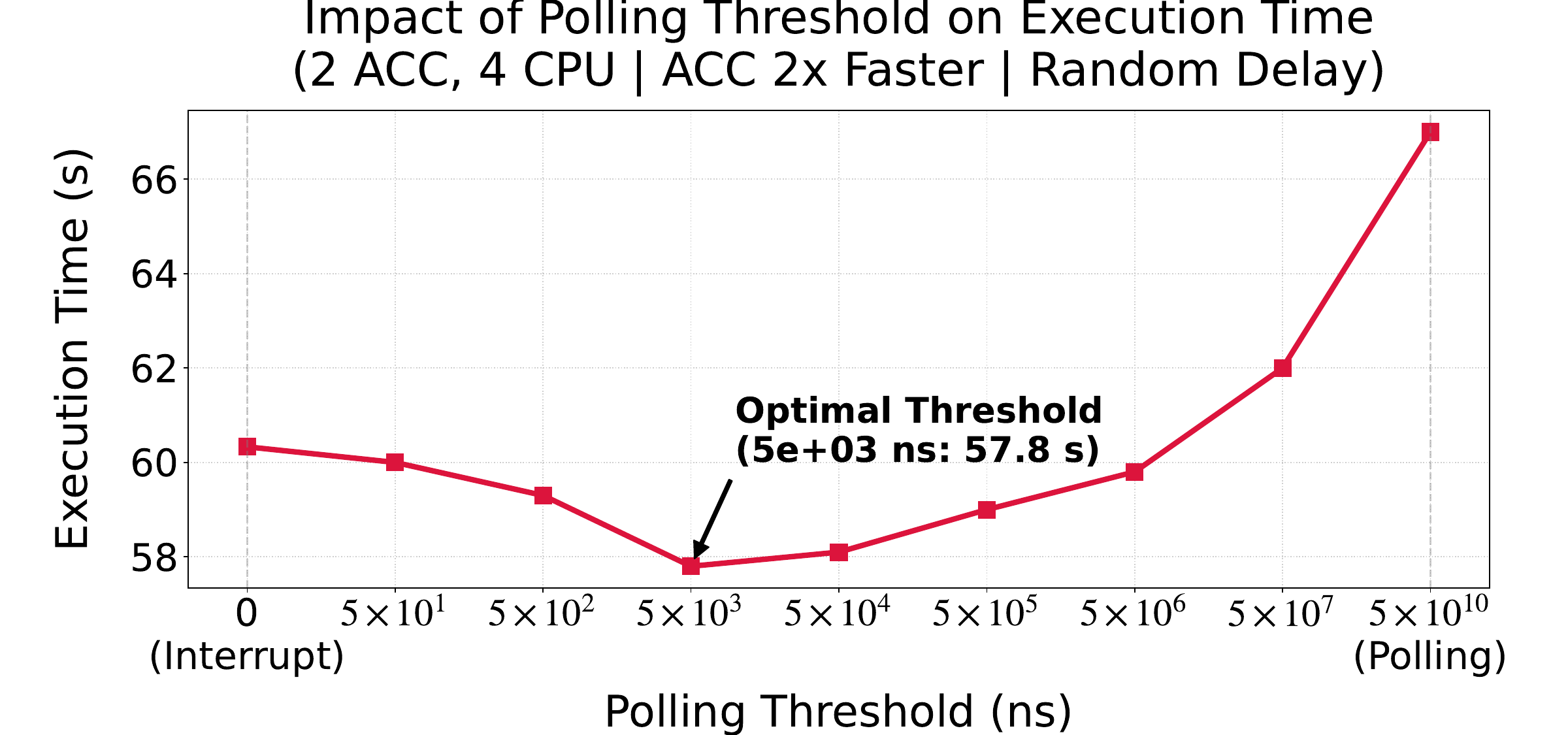}
    \caption{Performance of the proposed IPA for different interrupt-to-polling transition thresholds in SA. 
    }
    \label{fig:IPA}
\end{figure}

To evaluate IPA, we constructed a benchmark containing both short- and long-running accelerator tasks, forcing the scheduler to make completion-policy decisions across a wide range of execution times. Figure~\ref{fig:IPA} compares several threshold values used to switch between polling and interrupts. The results show that a threshold of 5000~ns consistently provides the best performance. This value closely matches the measured context-switch overhead of our interrupt-handling path, indicating that the optimal transition point occurs when the expected accelerator execution time becomes comparable to the operating-system cost of servicing an interrupt.

\section{Conclusion and Future Works}
\label{sec}
This paper demonstrates that treating accelerators as first-class CE rather than CPU-managed peripherals can substantially reduce the SW taxes that limit heterogeneous system performance. Using our FPGA-based research framework, we identified four major sources of overhead---communication, memory translation, interrupt handling, and scheduling---and proposed HW-SW co-designed mechanisms that reduce these costs, achieving up to $260\times$ lower communication overhead and improving application wall clock time by $2$--$6\times$ with a better scheduler.

Future work will further increase accelerator autonomy by offloading virtual-memory translation through hardware IOMMU support, introducing streaming DMA to overlap communication with computation, exploring hardware-assisted scheduling, and extending the framework to dynamically reconfigurable heterogeneous systems. Together, these directions move toward heterogeneous platforms where CPUs and accelerators operate as equal, composable computing resources.


\bibliographystyle{IEEEtranS}
\bibliography{refs}

\end{document}